\documentclass[12pt,longbibliography]{amsart}
\usepackage{geometry}
\usepackage{graphicx}
\usepackage{amsmath,amssymb,amsfonts}
\usepackage{amsthm}
\usepackage{mathrsfs}
\usepackage[title]{appendix}
\usepackage{xcolor}
\usepackage{algorithm}
\usepackage{algorithmicx}
\usepackage{algpseudocode}
\usepackage{listings}
\usepackage{enumitem}
\usepackage{nicematrix}
\usepackage{xspace}
\usepackage{hyperref}

\newtheorem{theorem}{Theorem}[section]

\newtheorem{observation}[theorem]{Observation}

\theoremstyle{definition}
\newtheorem{definition}[theorem]{Definition}
\newtheorem{remark}[theorem]{Remark}
\newtheorem{example}[theorem]{Example}

\DeclareMathOperator{\rank}{rank}
\newcommand{\cS}{{\mathcal S}}
\newcommand{\cA}{{\mathcal A}}
\newcommand{\RR}{{\mathbb R}}
\newcommand{\FF}{{\mathbb F}}
\newcommand{\row}{\operatorname{row}}
\newcommand{\col}{\operatorname{col}}
\newcommand{\MATLAB}{\textsc{Matlab}\xspace}

\newcommand{\Addresses}{{
  \bigskip
  \footnotesize
  R.~Ibrahim, \textsc{Department of Mathematics, Lafayette College, Easton, PA 18042 }
  \\
  \textit{E-mail address}, R.~Ibrahim: \texttt{ibrahimr@lafayette.edu}
  
  \medskip
 
 A.~H.~Moore, \textsc{Department of Mathematics \& Applied Mathematics, Virginia Commonwealth University, Richmond, VA 23284}
\\
  \textit{E-mail address}, A.~H.~Moore: \texttt{moorea14@vcu.edu}
}}

\begin{document}

\title[RNA Pseudoknots, Chord Diagrams and Intersection Graphs]{Methods for Analyzing RNA Pseudoknots via Chord Diagrams and Intersection Graphs}
 
\author{
Rayan Ibrahim 
\and 
Allison H. Moore
}



\maketitle

\begin{abstract}
RNA molecules are known to form complex secondary structures including pseudoknots.
A systematic framework for the enumeration, classification and prediction of secondary structures is critical to determine the biological significance of the molecular configurations of RNA.  
Chord diagrams are mathematical objects widely used to represent RNA secondary structures and to analyze structural motifs, however a mathematically rigorous enumeration of pseudoknots remains a challenge. 
We introduce a method that incorporates a distance-based metric $\tau$ to analyze the intersection graph of a chord diagram associated with a pseudoknotted structure.
In particular, our method formally defines a pseudoknot in terms of a weighted vertex cover of a certain intersection graph constructed from a partition of the chord diagram representing the nucleotide sequence of the RNA molecule. 
In this graph-theoretic context, we introduce a rigorous algorithm that enumerates pseudoknots, classifies secondary structures, and is sensitive to three-dimensional topological features. 
We implement our methods in MATLAB and test the algorithm on pseudoknotted structures from the bpRNA-1m database. 
Our findings confirm that genus is a robust quantifier of pseudoknot complexity. 
\end{abstract}

\section{Introduction \label{sec:intro} }

Ribonucleic acid (RNA) is a molecule essential to many functions of life, notably gene expression, cellular communication, and the storage and transfer of genetic information. The primary structure of RNA refers to the sequence of its four nitrogenous bases  adenine (A), guanine (G), cytosine (C), and uracil (U), attached along a sugar-phosphate backbone \cite{MolecularBiologyofRNA}. It is well known that RNA molecules fold into a variety of secondary and tertiary structures related to their natural functions via complementary Watson-Crick base pairings and other pairings \cite{MolecularBiologyofGene,MolecularBiologyofRNA}. 
Common secondary structure motifs include hairpin loops, stems (i.e. `stacks'), bulges,  interior loops, multiloops, single-stranded regions, and pseudoknots \cite{Reidys2011book}, \cite[Figure~1]{bpRNA} (see Figure \ref{fig:secondary_structures}). 
A pseudoknot is a secondary structure motif representing a three-dimensional folding pattern.
Pseudoknots were first recognized in the study of the turnip yellow mosaic virus \cite{Rietveld1982,Dumas1987}, but the term was coined in \cite{Studnicka1978}. The simplest type of pseudoknotted structure is an H-type pseudoknot, formed when nucleobases along the loop of a hairpin bond with nucleobases elsewhere along the sequence \cite{Brierley2007}. A variety of pseudoknot motifs have been characterized including H-type, K-type, L-type, and M-type motifs \cite{Kucharik2015,Andrikos_2022}.

RNA secondary structures may be represented graphically with  \emph{chord diagrams}, objects common in enumerative combinatorics and topology. 
In the representation of an RNA secondary structure, bonded pairs are indicated by arcs (`chords') along a line segment or circle. 
A precursor to a chord diagram representing secondary structures appears in \cite{Waterman2_1978} as a connection to predictive models using base-pairing matrices \cite{Tinoco1971}. There, RNA secondary structures were defined as simple planar graphs on a set of $n$ labeled points such that a path along the $n$ points represents the primary structure, with other edges representing bonds between bases \cite[Definition~2.1]{Waterman2_1978}. These planar graphs correspond to crossingless chord diagrams, which have since been studied extensively as models for secondary structures \cite{BON2008,Reidys2011,Waterman1978,Waterman1979,Penner1993,Waterman1994}. 
Pseudoknots occur only when the corresponding chord diagrams contain chords that cross each other.
Despite the relative simplicity of a chord diagram, there is no greed-upon method for quantifying the complexity of RNA pseudoknotting. A naive count of crossings overemphasizes contributions from helical stacking, and different methods for reducing parallel bonds may yield different enumerations of pseudoknots. Moreover, existing methods may ignore some topological features of the 3D conformation. We investigate these discrepancies in Sections \ref{subsec:pkstructures} and \ref{subsec:taureduction}. 

Our goal is to construct a mathematically rigorous and topologically robust framework for quantifying pseudoknot complexity in RNA, presented in the familiar language of graph theory and building upon conventions implicitly assumed in the bpRNA method \cite{bpRNA} and bpRNA-1m database \cite{bpRNAwebsite}. The bpRNA-1m database aggregates over 100K 
RNA secondary structures from seven sources \cite{CRW,tmRDB,SRPDB,tRNAdb,ribonucleaseP,Rfam,RCSBPDB}. Internal annotation routines and the enumeration of pseudoknots and other structural motifs are conducted via the algorithmic tool bpRNA \cite{bpRNA}. 
This database provides the primary test case for our graph-theoretic procedures. The strategy presented here verifies the reproducibility of both our methods and those of the bpRNA tool \cite{bpRNA}, while illustrating discrepancies resulting from topological conformations.

Section \ref{sec:theory} develops a mathematical formulation of pseudoknotting using chord diagrams and intersection graphs. The relevant graph theoretical background is reviewed in Section \ref{subsec:chord}, and we prove a relationship between vertex cover numbers and the genus of a chord diagram in Theorem \ref{thm:genusvertexcover}. 
In Section \ref{subsec:pkstructures}, we investigate precedent theories of pseudoknotting and convey these notions into our mathematical framework. 
Specifically, we use a weighted vertex cover of an intersection graph constructed from a partition of the chord diagram corresponding to an RNA molecule to give a precise enumeration of pseudoknots. 
In Section \ref{sec:methods}, we introduce the \emph{$\tau$-reduction algorithm}, which systematically reduces the complexity of a chord diagram to quantify pseudoknots in a robust manner that takes into account 3D topological features, including nestings, crossings, and a distance-based threshold $\tau$.
An explicit MATLAB implementation of our algorithm is provided in Github \cite{github}. 
Examples \ref{ex:weightedexample} and \ref{ex:similar} in this section demonstrate discrepancies in the existing methodology that are corrected by our current methods. 
In Section \ref{sec:discussion} we present the results of our analyses. 
We report quantities of pseudoknots, improving upon previous methods. 
In \cite{BON2008} it was shown that these basic motifs constitute the irreducible pseudoknots of genus equal to one (see Section \ref{sec:genus} for a discussion on genus). Moreover, they posit that the topological genus of a chord diagram provides a classification of RNA secondary structures
with pseudoknots. 
Our analysis in Section \ref{sec:genus} together with Theorem \ref{thm:genusvertexcover} confirms that even with additional topological considerations, genus is a robust classifier of pseudoknot complexity.

\section{Combinatorial Theory}
\label{sec:theory}


\subsection{Chord Diagrams}
\label{subsec:chord}

A \textit{linear chord diagram} $D$ is a set of $n$ points on an oriented line segment together with a (partial) matching of the points.
Circular chord diagrams are obtained by joining the endpoints of the segment; however we will restrict to linear chord diagrams throughout. 
We denote chords as pairs $c = (\ell, r)$, where $\ell$ and $r$ denote left and right endpoints, respectively. Because chord diagrams are matchings, $\ell < r$, and no two chords share an endpoint. By convention, a set of chords is indexed by left endpoints.

\begin{definition}\label{def:nesting-crossing}
For any two chords $c_1$ and $c_2$, there are three possibilities:
\begin{enumerate}[label=(\alph*)]
    \item $c_1$ and $c_2$ form a \textit{crossing}: $\ell_1 < \ell_2 < r_1 < r_2$.
    \item $c_1$ and $c_2$ form a \textit{nesting}: $\ell_1 < \ell_2 < r_2 < r_1$.
    \item $c_1$ and $c_2$ are \textit{independent}: $\ell_1 < r_1 < \ell_2 < r_2$.
\end{enumerate}    
\end{definition}

Further, a $k$-crossing is a set of chords $(\ell_1,r_1),(\ell_2,r_2),\dots, (\ell_k,r_k)$ such that $\ell_1 < \ell_2 < \cdots < \ell_k < r_1 < r_2 < \cdots < r_k$. A $k$-nesting is a set of chords $(\ell_1,r_1),(\ell_2,r_2),\dots, (\ell_k,r_k)$ such that $\ell_1 < \ell_2 < \cdots < \ell_k < r_k < r_{k-1} < \cdots < r_1$.

\begin{figure}[ht]
\centering
\includegraphics[width=0.6\textwidth]{crossing_nesting_ex.pdf}
\caption{Top left: A $3$-nesting. Top right: A $3$-crossing. Bottom left: Two $2$-crossings. Bottom right: A $2$-nesting and two $2$-crossings.}
\end{figure}

The following definitions will become useful in formalizing pseudoknots in chord diagrams. We use the notation $(a,b)$ for the open interval between $a$ and $b$. 

\begin{definition}[Chord Obstructed]\label{def:chordobstructed}
    Let $c$ and $c'$ be chords and let $U = (\ell,\ell') \cup  (r,r')$, i.e. $U$ can be thought of as the set of bases between the left endpoints and right endpoints of $c$ and $c'$. We say $c$ and $c'$ are \textit{chord obstructed} if there is some chord $c''$ such that either $\ell'' \in U$ or $r''\in U$. A set of three or more chords $S = \{c_1,c_2,\dots,c_k\}$ is chord obstructed if consecutive chords $c_i$ and $c_{i+1}$ are chord obstructed for some $i = 1,2,\dots,k-1$.
\end{definition}

\begin{definition}[Segment]\label{def:segment}
    A \textit{segment} of a chord diagram $D$ is a maximal nonempty set of chords $S = \{c_1,c_2,\dots,c_k\}$ forming a $k$-nesting such that $S$ is not chord obstructed.
\end{definition}
Note that the set of segments $\cS$ partitions the set of chords $C$. In crossingless chord diagrams, there is a natural poset structure on the set of segments defined by $S \prec S'$ if $S'$ is nested in $S$.

The \textit{intersection graph} $G$ of a chord diagram $D$ is the graph whose vertices are the chords of $D$, and such that two vertices in $G$ are adjacent if their corresponding chords in $D$ form a crossing.
The use of intersection graphs is well established, see for example \cite{McKee1999}.
Variations on the concept of an intersection graph arise numerous times in the biology literature under different names. For example, in \cite{Kucharik2015}, there is the notion of a conflict graph, whose vertex set comprises helices in the RNA structure and edges signify the crossing of chords corresponding to the helices. In \cite{Shu2008} the concept of an element-contact graph is introduced, in particular the stem-loop-contact graph (SLCG) \cite[Figure 7]{Shu2008}. The segment graph of the bpRNA database \cite{bpRNA} is another such example. 

We will investigate the following graph theoretic invariants with respect to RNA structures in Section \ref{sec:discussion}. An \emph{independent set} is a set of vertices such that no two vertices in the set are adjacent. 
The maximum \textit{clique number} $\omega(G)$ is the maximum size of a complete subgraph of $G$. 
    A \textit{vertex cover} of a graph $G$ is a set $A\subseteq V(G)$ such that for every edge $xy \in E(G)$ either $x\in A$ or $y\in A$. The \textit{vertex cover number} of a graph $G$, denoted $\beta(G)$, is the number of vertices in a minimum vertex cover.
    The \textit{weight} of a vertex cover $A$ in a vertex-weighted graph $G$ is $\sum_{v\in A} w(v)$. 
Note that a \textit{minimum weight vertex cover} of a weighted graph $G$ is not necessarily a minimum cardinality vertex cover. (Consider for example the path graph $P_3$ with weights 1,5,3). 

If $C$ is a vertex cover of $G$, then the graph $G-C$ contains no edges, as by definition every edge of $G$ must have an endpoint in $C$. Similarly, if $I$ is an independent set of $G$, then every edge of $G$ has at least one end point in $G-I$. Thus we have the following observation.

\begin{observation}\label{obs:indset_cover}
    Let $G$ be a graph and let $I$ and $C$ be an independent set and vertex cover respectively. Then $V(G)\setminus I$ is a vertex cover, and $V(G) \setminus C$ is an independent set. Moreover, the complement of a minimum weight vertex cover is a maximum weight independent set.
\end{observation}

\begin{definition}[Genus]\label{def:genus}
    The \textit{genus} of a chord diagram $D$, denoted $\gamma(D)$, is half of the rank of the adjacency matrix of the intersection graph with $\mathbf{Z}_2$ coefficients \cite{Moran1984}.
\end{definition}

\begin{figure}[ht]
    \centering
    \includegraphics[width=0.8\textwidth]{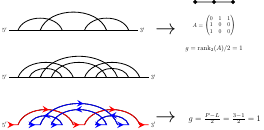}
    \caption{Two methods of calculating the genus of a given chord diagram.}
    \label{fig:genus_calc}
\end{figure}

Equivalently, the genus of a chord diagram is equal to the topological genus of the surface obtained by regarding the chord diagram as a band surgery diagram \cite{BON2008}. One way to calculate the genus $g$ of a chord diagram $D$ is via the formula \[g = \frac{P-L}{2}\] where $P$ is the number of chords (or base pairs) and $L$ is the number of closed loops in the corresponding double-line diagram \cite{BON2008} (see Figure \ref{fig:genus_calc}.)
For a graph $G$ we define $\rank(G)$ (resp. $\rank_p(G))$ to be the rank of the adjacency matrix of $G$ with coefficients in $\RR$ (resp. coefficients in $\FF_p$).

\begin{theorem}\label{thm:genusvertexcover}
    Let $D$ be a chord diagram and let $G$ be the intersection graph of $D$. If $G$ is acyclic, then $\gamma(D) = \beta(G)$.
\end{theorem}

\begin{proof}
    Because $\beta(G)$ and $\gamma(G)$ are additive over disjoint unions, without loss of generality we may assume $G$ is a tree of order $n$.
    The statement is easily verified for $n \le 4$. 
    We will use induction on $n$.
    Let $T$ be a tree on $n\ge 5$ vertices with adjacency matrix $A$. If $T$ has a vertex $v$ with at least two leaf neighbors, then we remove a leaf $\ell$ adjacent to $v$ to form $T'$. By the induction hypothesis, $\rank_2(T')/2 = \beta(T')$. Adding back $\ell$ to form $T$, we have $\rank_2(T') = \rank_2(T)$ as the row corresponding to $\ell$ in $A$ is identical to the rows corresponding to the other leaf neighbors of $v$. Note in a graph, for any vertex with a leaf neighbor, there is a minimum vertex cover containing that vertex. Thus, we have $\beta(T) = \beta(T')$.

    Now we assume every vertex of $T$ has at most one leaf as a neighbor. Then there is some leaf $\ell$ with a neighbor $v$ of degree two; to identify such a vertex $v$, find a maximum path and let $v$ be adjacent to a leaf. Let $\{\ell,w\} = N(v)$. Let $T' = T - \{v,\ell\}$. By the induction hypothesis, $\gamma(T') = \beta(T')$. 
    
    First we claim $\beta(T) = \beta(T')+1$. Indeed, no minimum vertex cover of $T'$ will cover the edge $v\ell$, so $T$ requires one more vertex in addition to a minimum vertex cover of $T'$ to cover all edges of $T$. 

Next we claim that $\rank_2(T) = \rank_2(T') + 2$. The adjacency matrix of $T$ is given by 
\[A = 
\setlength{\arraycolsep}{5pt}
\renewcommand\arraystretch{0.65}
\begin{pNiceArray}{c c c c | c c}[first-col,first-row]
& &&& w & v & \ell \\
& \Block{3-3}<\Large>{} &&&\ast & 0 & 0\\ 
&                         &&&\vdots & \vdots & \vdots  \\ 
&                         &&&\ast & 0 & 0\\
w &\ast & \cdots & \ast & 0 & 1 & 0 \\ \hline
v &0 & \cdots & 0 & 1 & 0 & 1\\
\ell &0 & \cdots & 0 & 0 & 1 & 0 
\end{pNiceArray}
\]
where the upper left block corresponds to the adjacency matrix $A'$ of $T'$. Applying the row operation $\operatorname{row} w - \row \ell = \row v$ and column operation $\col w - \col \ell = \col w$ we obtain the matrix 
\[ B=
\setlength{\arraycolsep}{5pt}
\renewcommand\arraystretch{0.65}
\begin{pNiceArray}{c c c c | c c}[first-col,first-row]
& &&& w & v & \ell \\
& \Block{3-3}<\Large>{} &&&\ast & 0 & 0\\ 
&                         &&&\vdots & \vdots & \vdots  \\ 
&                         &&&\ast & 0 & 0\\
w &\ast & \cdots & \ast & 0 & 0 & 0 \\ \hline
v &0 & \cdots & 0 & 0 & 0 & 1\\
\ell &0 & \cdots & 0 & 0 & 1 & 0 
\end{pNiceArray}.
\]
We have that $\rank_2(A) = \rank_2(B) = \rank_2(A')+2$, therefore $\gamma(T)=\gamma(T')+1$.
This completes the induction. 
\end{proof}

\begin{remark}
    In general, for acyclic graphs $\gamma(G)\neq \beta(G)$. In particular, for complete graphs, one may observe that $\beta(K_n)-\gamma(K_n)=k - \left \lfloor{ \frac{k}{2} }\right \rfloor  -1 $.
\end{remark}

Finally, we remark that the genus of a chord diagram $D$ containing an $r$-nesting $C=\{c_1, \cdots, c_r\}$ that is not chord obstructed is equal to the genus of the diagram with $r-1$ chords of $C$ removed, i.e. $D - \{c_2, \cdots, c_r\}$. The rank of a matrix can be thought of as the maximum number of linearly independent rows or columns in the matrix. It can be seen from the adjacency matrix $A$ of the intersection graph that $c_2,\dots,c_r$ correspond to identical rows in $A$, and are thus linearly dependent.

\subsection{Pseudoknotted Structures}\label{subsec:pkstructures}

Analyzing the role of pseudoknots in various RNA processes motivates the study of pseudoknot complexity for comparison, characterization of motifs, and prediction of RNA secondary structure \cite{bpRNA,BON2008,Reidys2011,Staple2005}. Practical definitions of a pseudoknot vary in the biological literature \cite{BON2008,Reidys2011}. Because we will ultimately be interested in quantifying pseudoknots aggregated by the bpRNA-1m database \cite{bpRNA}, we start from definitions presented there. In this database, a pseudoknotted structure is characterized as having base pair positions that cross in the sense of Definition \ref{def:nesting-crossing}(a); the working definition of a `pseudoknot base pair’ is one belonging to a minimal set of base pairs that results in a pseudoknot-free structure once removed. There is some ambiguity in these concepts; for example, a pair of kissing hairpins with intersection graph weighted $[a, a+b, b]$ demonstrates that such a minimal set is not unique.  
Further ambiguities may result from helices formed in secondary structures (see Section \ref{sec:similar_pseudoknotted_structures}). 
By default, the number of pseudoknots in any given secondary structure reported in \cite{bpRNA} is the number returned by an algorithm that identifies some minimal set of pseudoknot base pairs. 
We review their algorithm, and translate corresponding notions of pseudoknotting into the language of chord diagrams, as follows.

To make the notion of a pseudoknot, and more specifically the annotation of multiple pseudoknots, more precise, \cite{bpRNA} introduces the notion of a segment of RNA secondary structure. An RNA segment is described as a region of duplexed RNA, possibly containing bulges or internal loops. In combinatorial terms, RNA segments correspond to the segments $s\in \cS$ of the chord diagram $D$ representing the secondary structure, as in Definition \ref{def:segment}. The segments partition the set of chords $C$, and ordering the base sequence from the 5'-end to 3'-end indexes each segment by its leftmost endpoint. 
In \cite{bpRNA}, the following is observed; we provide a restatement in terms of chord diagrams.
\begin{theorem}[\cite{bpRNA}]\label{thm:segment}
    Let $\cS$ be the segment partition of a linear chord diagram $D$ and let $S,S' \in \cS$. If there are chords $c \in S$ and $c' \in S'$ such that $c$ and $c'$ are crossed, then any pair of chords from $S$ and $S'$ cross. 
\end{theorem}

\begin{proof}
Recall that a segment of size $k$ is a maximal $k$-nesting in which pairs of consecutive chords are not chord obstructed. Let $\cS$ be a segment partition of $D$ and let $S,S' \in \cS$ where $S < S'$ in the indexing of segments by left endpoints. Let $c \in S$ and $c'\in S'$ such that $c$ and $c'$ cross. Let $\ell_{\max}$ be the maximum left endpoint of $S$ and $r_{\min}$ and $r_{\max}$ be the minimum and maximum right endpoints of $S$ respectively. 

By the left endpoint indexing, all left endpoints of chords in $S'$ must be greater than $\ell_{\max}$. If some left endpoints of $S'$ are less than $r_{\min}$ and some greater, then $S'$ is chord obstructed by the right endpoints of $S$. If all left endpoints of $S$ are greater than $r_{\min}$, either $S$ is chord obstructed or no chords of $S$ and $S'$ cross. Thus, all left endpoints of $S'$ must lie between $\ell_{\min}$ and $r_{\min}$. Since $c$ and $c'$ cross, and $S$ is not chord obstructed, it must be that $r' > r_{\max}$, i.e. $c'$ crosses every chord in $S$. No other right endpoint of $S'$ is less than $r_{\max}$, as otherwise $S$ or $S'$ are chord obstructed. Thus any pair of chords from $S$ and $S'$ cross.\end{proof}

In other words, Theorem \ref{thm:segment} says that two segments $s,s'\in S$ cross whenever any chords $c\in s, c'\in s'$ cross. The \textit{segment graph} $G_\cS$ of a chord diagram is a variation on an intersection graph whose vertex set is the set of segments, where two vertices are adjacent if their segments cross \cite{bpRNA}. Vertices in $G_\cS$ are weighted by the number of chords contained in their corresponding segments. 
The terminology \textit{PK-segment} refers to any segment $s\in\cS$ that crosses any other segment. Let $H_\cS\subset G_\cS$ be the subgraph with isolated vertices removed, referred to as the \textit{PK-segment graph} in \cite{bpRNA}.  

In \cite{bpRNA}, pseudoknotted structures are identified by finding a maximum weight independent set $I$ in $H_\cS$ via a heuristic approach, with an exact algorithm used in the specific case of components which are paths. As we have formalized above, the set $P = V(G_\cS)-I$ is a minimum weight vertex cover of $G_\cS$. That is, $P$ is a set of segments of minimum cardinality in $C$ that when removed from $D$ leave a pseudoknot-free structure. 

We may now formally quantify the size of a pseudoknotted structure according to the conventions of the bpRNA-1m database, as implied by the algorithms of \cite{bpRNA}.

\begin{definition}(Pseudoknotted Structures - \textit{bpRNA Segment Graph Method})
\label{def:pseudoknot-segment}
 A secondary structure is \textit{pseudoknotted} if its segment graph contains at least one edge  and is called \textit{pseudoknot-free} otherwise. The \textit{number of pseudoknots} in a pseudoknotted structure is the minimum cardinality over all vertex covers of minimum weight of the segment graph.  
\end{definition}
Any segment contained in a minimum-cardinality minimum-weight vertex cover may be called simply `a pseudoknot.' 
It is important to note that in \cite{bpRNA}, the number of pseudoknots is not the number of chords in the corresponding cover, in general.

\begin{figure}[ht]
    \centering
    \includegraphics{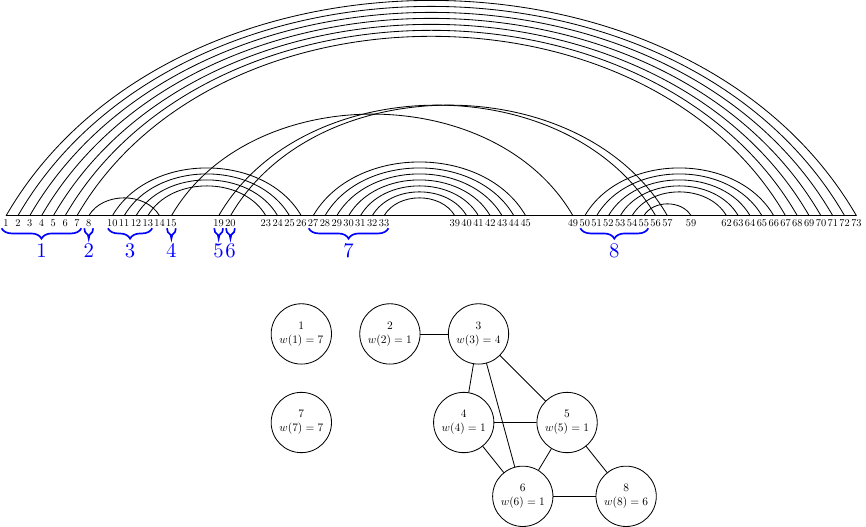}
    \caption{(Top) Linear chord diagram of transfer RNA molecule of type 76-MER from \textit{Escherichia coli} (PDB\_652) with segments labeled. (Bottom) Segment graph $G_\cS$ with weights. }
    \label{fig:pdb652segments}
\end{figure}

\begin{example}
Figure \ref{fig:pdb652segments} shows the chord diagram representing the secondary structure of tRNA (76-MER) found in \textit{Escherichia coli} with 8 segments. The PK-segments are $\{2,3,4,5,6,8\}$. The maximum weight independent set in $H_\cS$ is $\{3,8\}$. The minimum weight vertex cover in $G_\cS$ is $\{2,4,5,6\}$, indicating four pseudoknots in this structure according to the conventions of \cite{bpRNA}. 
\end{example}

\subsection{Crossingless Secondary Structures}
\label{subsec:secondary_structures}
\begin{figure}[ht]
    \centering
    \includegraphics[width=\textwidth]{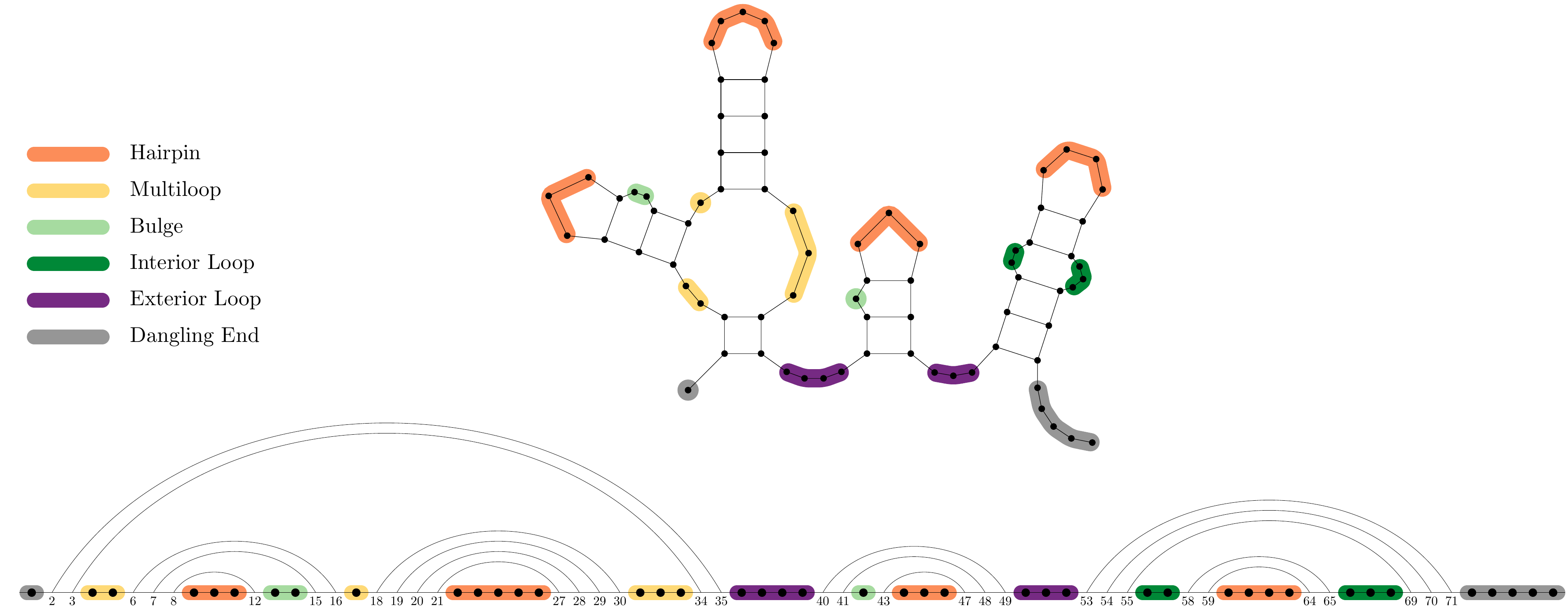}
    \caption{An illustrated example of secondary structures in a crossingless chord diagram. There are eight stems in this structure}
    \label{fig:secondary_structures}
\end{figure}

    Let $D$ be a chord diagram with no crossings, and let $\cS$ be the segment partition of $D$. We say an unpaired base is nested in a segment $S$ if it is between the left and right endpoints of the innermost chord of $S$. A \textit{stem} in $D$ is a $k$-nesting with no other bases between the endpoints of any two consecutive chords. Segments are composed of stems. Let $S$ be a segment and let $c$ and $c'$ be two consecutive chords in a segment $S$, with $\ell < \ell'$. If there are sequences of unpaired bases in the intervals $(\ell,\ell')$ and $(r,r')$, i.e. $\ell' - \ell \ge 2$ and $r-r' \ge 2$, then those two sequences together comprise an \textit{interior loop}. If exactly one of $(\ell,\ell')$ or $(r,r')$ contains a sequence of unpaired bases, that sequence is a \textit{bulge}. If two or more segments $\mathcal{T}$ are nested in $S$, then there is a \textit{multiloop} composed of all unpaired bases $b$ nested in $S$ and not nested in any segment in $\mathcal{T}$. Note multiloops may have length zero (that is, a multiloop may be the empty set). The \textit{exterior loop} of $D$ is the set of all unpaired bases which are not nested in any chord. If $b_0$ is the first paired base and $b_f$ is the last paired base in the base sequence, the set of unpaired bases less than $b_0$ and greater than $b_f$ are a part of the exterior loop called the \textit{dangling ends}. The exterior loop can be thought of as the multiloop gained from an imaginary base pair bonding of the $5'$-end and $3'$-end, under which all chords are nested. The crossingless secondary structures are illustrated in Figure \ref{fig:secondary_structures}.

\subsection{Qualitatively Similar Pseudoknotted Structures}
\label{sec:similar_pseudoknotted_structures}

Identifying nested chords in a chord diagram is a common strategy for reducing the complexity of the combinatorial analysis of secondary structures because it allows for the reduction of helices to a single chord, or alternatively to a single vertex in an intersection graph. For example, nested chords are identified as segments in \cite{bpRNA} (Definition \ref{def:segment} above), as `stacks' in \cite{BON2008}, as `shadows' in \cite{Reidys2011}, or as edges in tree diagrams of pseudoknot-free structures in \cite{Benedetti1996} and elsewhere. Such reductions also preserve some invariants of interest, for example the topological genus (see Definition \ref{def:genus} and \cite{BON2008}). In contrast to $r$-nestings, $r$-crossings in chord diagrams are typically left untouched by simplification algorithms. Consequently, the existence of an $r$-crossing implies the existence of at least $r$ pseudoknots according to the conventions of \cite{bpRNA}.

Consider the following two examples.

\begin{example}[Discrepancies due to weights]
\label{ex:weightedexample}
Consider two weighted segment graphs, each isomorphic to $P_3$, with weights $(1,5,3)$ and $(1,5,4)$, respectively.  Such graphs represent nearly identical $K$-type secondary structures which differ only by a single bonded pair in the third stem. By \cite{bpRNA} and Definition \ref{def:pseudoknot-segment}, the minimum cardinality over vertex covers of minimum weight determines the number of pseudoknots: 2 and 1, respectively. Note that the addition of one pair results in a \textit{decrease} in the number of pseudoknots.
\end{example}

\begin{figure}[ht]
\centering
\includegraphics[width=\textwidth]{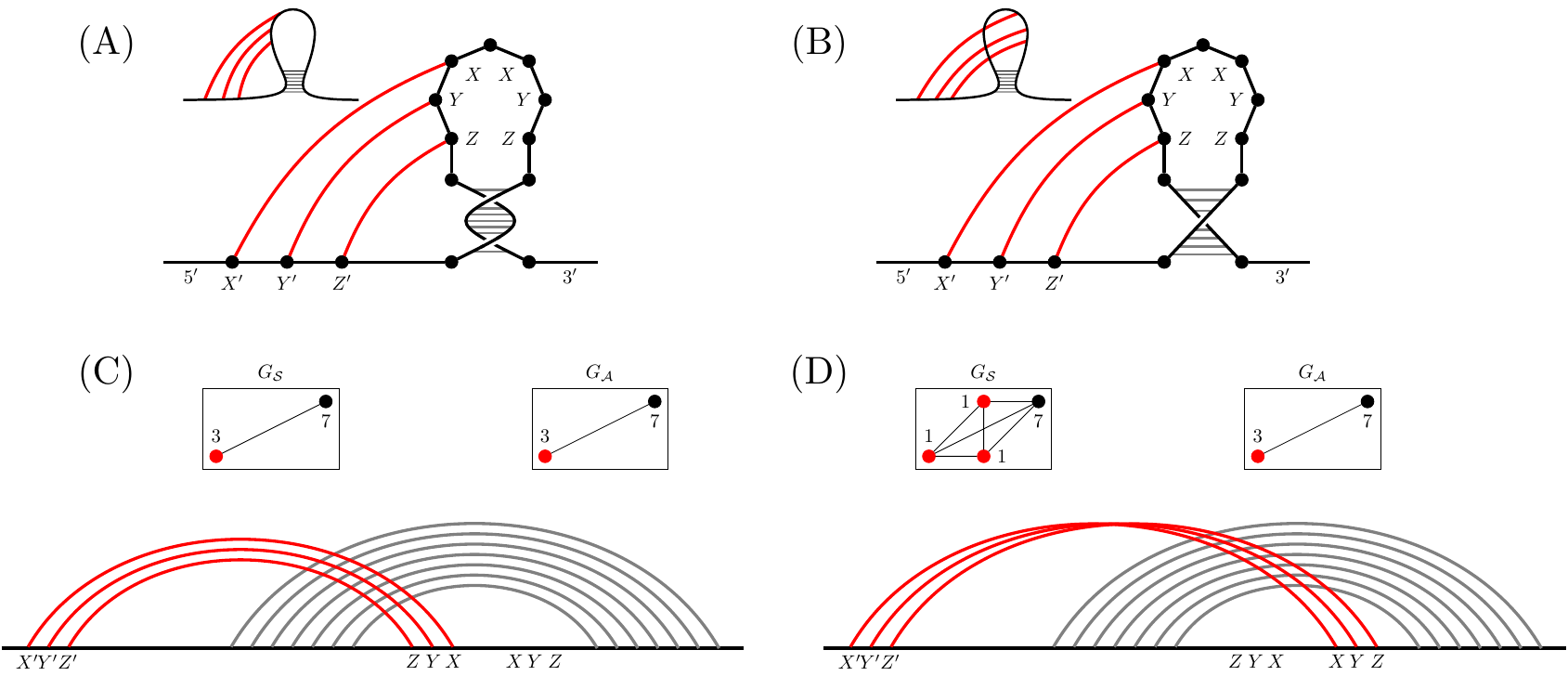}
\caption{Two closely related instances of bonding between two hairpins} \label{fig:inversion_example}
\end{figure}

\begin{example}[Discrepancies due to $r$-crossings]
\label{ex:similar}
Consider a nucleotide sequence that contains a repetitious subsequence 
appearing in reverse. 
For an example, we follow Figure \ref{fig:inversion_example}.
This structure contains subsequence $\sigma=X'Y'Z'$, complementary sequence $\sigma'=ZYX$, and reverse complementary sequence $\overline{\sigma}'=XYZ$. Bonds formed between $\sigma$ and $\sigma'$ result in an $r$-nesting (Figure \ref{fig:inversion_example}(A)), whereas bonds formed between $\sigma$ and $\overline{\sigma}'$ form an $r$-crossing (Figure \ref{fig:inversion_example}(B)). We assume here that either set of bonds is possible. In particular, the parity of number of half-turns in the helical stem may determine whether $\sigma'$ or $\overline{\sigma}'$ is nearer to $\sigma$ in a 3D conformation. Despite the heuristic similarity of the resulting pseudoknotted structures, their segment graphs differ significantly.
By the conventions of \cite{bpRNA} (Definition \ref{def:pseudoknot-segment}) the two structures are of pseudoknotting size 1 and 3, respectively. Moreover, the difference increases with additional base pairing in the hairpin stem.
\end{example}

The general observation is that variances in bonding from spatial conformations (e.g. helical twisting) may result in a quantification of pseudoknotting that is artificially high. Of $30$ structures exhibiting the most pseudoknotting in Table \ref{tab:mostPK}, there are $28$ that contain both a complementary and reverse complementary sequence (possibly not contiguous) near the site of the $3$-crossing. One example RNA structure involving a $3$-crossing and $5$-nesting is bpRNA\_CRW\_55315. This structure contains a $3$-crossing with left bases GCA at indices 2107, 2108, and 2112 and right bases AGU at indices 2164, 2165, and 2167, which is the reverse of the triple UGA at indices 2162, 2163, and 2164.

\section{Methods}
\label{sec:methods}

To further reduce complexity and to more effectively relate similar secondary structures, we propose in this section an alternative method for quantifying the size of a pseudoknotted structure and a new simplification algorithm that identifies both $r$-crossings and $r$-nestings in chord diagrams. In addition to handling complexity issues arising from $r$-crossings, this method will also eliminate some discrepancies resulting from weighted graphs and include a parameter accounting for distance between nucleotides.

\subsection{The \texorpdfstring{$\tau$}{tau}-Segment Graph Method}
\label{subsec:taureduction}

In this subsection, we give a partitioning procedure that identifies $r$-crossings in a manner similar to that of $r$-nestings. We implement the procedure in an algorithm that incorporates an additional distance parameter $\tau$ in terms of the nucleotide sequence.

     An \textit{augmented segment} of a chord diagram is a maximal nonempty set of chords $S = \{c_1,c_2,\dots,c_k\}$ forming a $k$-nesting or a $k$-crossing which is not chord obstructed. 
Revisiting Examples \ref{ex:weightedexample} and \ref{ex:similar} above, notice that an intersection graph $G_\cA$ produced with augmented segments would yield two pseudoknotted structures of size 1 in Example \ref{ex:weightedexample} and two structures of size 2 in Example \ref{ex:similar}. 

 The \textit{chord distance} between two chords $c_1 = (\ell_1,r_1)$ and $c_2 = (\ell_2,r_2)$ is \[d(c_1,c_2) = \max \{|\ell_1 - \ell_2|, |r_1 - r_2|\}.\]

We say that a pair of nested or crossed chords $c_1, c_2$ are \textit{$\tau$-near} if $d(c_1,c_2) \le \tau$ and $c_1$ and $c_2$ are not chord obstructed. A $k$-crossing or $k$-nesting $\{c_1,\dots,c_k\}$ is $\tau$-near if for each $i=1,\dots,k-1$ we have $c_i$ and $c_{i+1}$ are $\tau$-near.  
    A \textit{$\tau$-segment} of a chord diagram is a maximal nonempty set of chords $S = \{c_1,c_2,\dots,c_k\}$ forming a $\tau$-near $k$-nesting or a $\tau$-near $k$-crossing. 
As with segments or augmented segments, the set of $\tau$-segments $\cS_\tau$ also partitions the set of chords $C$. 
We have the following statement.

\begin{theorem}\label{thm:tau-segment}
    Let $\cS_\tau$ be the $\tau$-segment partition of a linear chord diagram $D$ and let $S_1,S_2 \in \cS_\tau$. If there are chords $c \in S_1$ and $c' \in S_2$ such that $c$ and $c'$ are crossed, then for every pair $c \in S_1$ and $c' \in S_2$, the chords $c$ and $c'$ are crossed. 
\end{theorem}

\begin{proof}
The proof is analogous to that of Theorem \ref{thm:segment}.
\end{proof}

Generalizing the segment graph, we may now define the $\tau$-segment intersection graph $G_\tau$ to be the 
    weighted graph whose vertex set is the set of $\tau$-segments, where two vertices are adjacent if the $\tau$-segments cross. When $\tau=0$, define $G_{\tau=0}:=G_\cS$. When $\tau=\infty$, the graph $G_\cA:=G_{\infty}$ is the intersection graph of the augmented segment partition. 
The notation $D_\tau$ will indicate the chord diagram corresponding to $G_\tau$, in which each $\tau$-segment corresponds to a chord. As with $G_\cS$ and $G_\cA$ we use the notation $D_\cS$ and $D_\cA$ analogously. We may now formally revise the method for quantifying the size of pseudoknotted structures in RNA.

\begin{definition}(Pseudoknotted Structures - \textit{$\tau$-Segment Graph Method})
\label{def:pseudoknot-tau}
 A secondary structure is $\tau$-\textit{pseudoknotted} if $G_\tau$ contains at least one edge and is called \textit{pseudoknot-free} otherwise. For $\tau\geq 1$, the \textit{number of pseudoknots} is the minimum cardinality of a vertex cover of $G_\tau$. For $\tau=0$, the number of pseudoknots is the minimum cardinality over all vertex covers of minimum weight of the segment graph.
\end{definition}
The definition in the case of $\tau=0$ is explicitly made to agree with the conventions of \cite{bpRNA}.

\begin{figure}
    \centering
    \includegraphics[width=\textwidth]{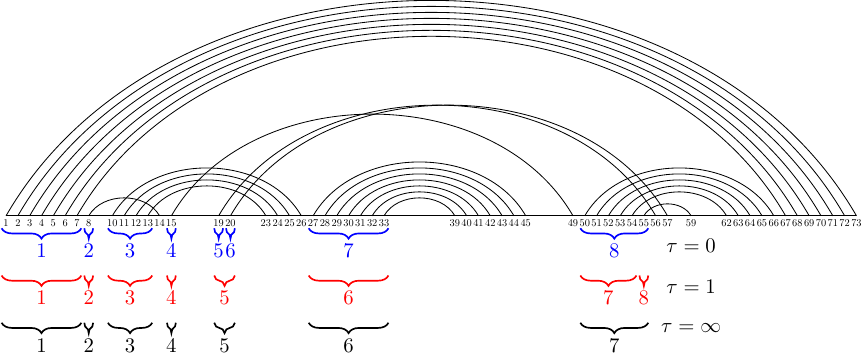}
    \caption{Different tau-partitions of bpRNA\_PDB\_652 for $\tau=0, 1$, and $\infty$.}
    \label{fig:tau0toInf}
\end{figure}

\begin{example}
    Figure \ref{fig:tau0toInf}. The maximum distance between any two non-chord obstructed chords is three. That is, the $\tau$-segment partition is the same for all $\tau \ge 3$. This RNA structure has four pseudoknots according to the conventions of \cite{bpRNA}, and two pseudoknots according to Definition \ref{def:pseudoknot-tau}. The main difference comes from chords $(19,56)$ and $(20,57)$ becoming part of the same segment for $\tau$ large enough. 
\end{example}

Algorithm \ref{alg:partitioner} in the next section implements the $\tau$-segment partition procedure.
\begin{algorithm}[ht]\caption{Tau-Segment Partition}\label{alg:partitioner}
\begin{algorithmic}
\Require $C = \{c_1,c_2,\dots,c_k\} = \{(\ell_1,r_1),(\ell_2,r_2),\dots, (\ell_k,r_k)\}$, $\tau$
\Ensure $\mathcal{C}_\tau$
\State $S \gets \{(\ell_1,r_1)\}$ 
\If{$\tau > 0$}
    \For{$1 \le i \le k-1$}
        \State $d \gets \max\{|\ell_i - \ell_{i+1}|, |r_i - r_{i+1}| \}$ \Comment{\small Distance.}
        \If{$d \le \tau$ \textbf{and} $ \neg \operatorname{isChordObstructed}(c_i,c_{i+1})$}
            \State $S \gets \operatorname{append}(S,(\ell_{i+1},r_{i+1}))$
        \Else
            \State $\mathcal{C}_\tau \gets \operatorname{append}(\mathcal{C}_\tau,S)$ \Comment{Store Segment}
            \State $S \gets \{c_{i+1}\}$ \Comment{Initialize new segment}
        \EndIf
    \EndFor
\Else
    \For{$1 \le i \le k-1$}
            \If{$\operatorname{isNested}(c_i,c_{i+1})$ \textbf{and} $\neg \operatorname{ChordObstructed}(c_i,c_{i+1})$}
                \State $S \gets \operatorname{append}(S,c_{i+1})$
            \Else
                \State $\mathcal{C}_\tau \gets \operatorname{append}(\mathcal{C}_\tau,S)$ \Comment{Store Segment}
                \State $S \gets \{c_{i+1}\}$ \Comment{Initialize new segment}
            \EndIf
    \EndFor
\EndIf
\State $\mathcal{C}_\tau \gets \operatorname{append}(\mathcal{C}_\tau,S)$ \Comment{Account for final segment.}
\end{algorithmic}
\end{algorithm}
The input to the algorithm is a chord diagram with set of chords $C$ (a list of base pairs indexed by left endpoint) and a non-negative integer parameter $\tau$. Selecting a pair of chords $c,c'\in C$, the algorithm loops to build the $\tau$-segment containing $c$. If $\tau=0$, it checks whether the $c,c'$ are nested and not chord obstructed. If $\tau>0$, the algorithm checks whether $c$ and $c'$ are $\tau$-near and not chord obstructed. If the criterion is met, $c'$ is added to the segment containing $c$ and the next pair is selected. If not, the segment is closed. The next unvisited pair of chords are then selected and the process repeats to build the next segment until all chords have been exhausted. 

The Github repository \cite{github} contains Algorithm \ref{alg:partitioner} implemented in \MATLAB in the function called \texttt{findSegments}.

\subsection{Pseudoknot Quantification Process}
\label{process}

Here, we apply the definitions and algorithm above to the bpRNA-1m(90) database \cite{bpRNA}. The database bpRNA-1m(90) is a subset of bpRNA-1m restricted to the 28,370 RNA secondary structures with less than 90\% sequence similarity. This database contains 3,320 RNA structures reported to contain at least one pseudoknot, i.e. structures whose segment graphs contain at least one edge, with a total of 7,164 pseudoknots reported according to Definition \ref{def:pseudoknot-segment} (the prior bpRNA Segment Graph Method). 
To analyze the data, we implement both the segment and $\tau$-segment graph methods and analyze secondary structures via \MATLAB code \cite{github}.

Chord diagrams associated with RNA structures are stored as \MATLAB arrays. The input to the Tau-Segment Partition algorithm is the set of chords $C$ from a chord diagram $D$ and an integer parameter $\tau$. To carry out any of the above methods, we first call the \texttt{findSegments} function to create a segment partition of the chord diagram. The parameter $\tau$ determines which segment partition is created, with $\tau=0,\infty, 1<\tau<\infty$ corresponding to the segment partition, augmented segment partition, and $\tau$-partition, respectively. See Figure \ref{fig:tau0toInf}.

Depending on whether the segment partition contains any segments which cross each other, one of two subroutines is implemented. If the segment partition contains no segments which cross, then the secondary structures of the chord diagram are analyzed by the function \texttt{classifyBases}. This function outputs the primary base sequence with each base classified as belonging to one of the secondary structure motifs given in Section \ref{subsec:secondary_structures}. 

If the segment partition contains segments which cross, then we construct the corresponding weighted segment graph with the function \texttt{makeSegmentGraph}.
The pseudoknots of the chord diagram are then identified and analyzed by the \texttt{findPKs} process:
\begin{enumerate}
    \item Find list of all maximal independent sets $I$.
    \item If $\tau = 0$:
    \begin{enumerate}
        \item Calculate weight of each set $I \in \mathcal{I}$.
        \item Subset all maximum weight sets $\mathcal{I}'\subseteq \mathcal{I}$.
        \item Subset maximum cardinality maximum weight sets $\mathcal{I}''\subseteq \mathcal{I}' \subseteq \mathcal{I}$.    
    \end{enumerate}
    \item If $\tau > 0$:
    \begin{enumerate}
        \item Subset all maximum cardinality sets $\mathcal{I}'\subseteq \mathcal{I}$.     
        \item Calculate weight of each set $I \in \mathcal{I}'$.
        \item Subset maximum weight maximum cardinality sets $\mathcal{I}''\subseteq \mathcal{I}' \subseteq \mathcal{I}$.    
    \end{enumerate}
    \item Dualize the independent sets $\mathcal{I}''$ to vertex covers $\mathcal{P}$.
    \item Select first vertex cover $P\in \mathcal{P}$ with respect to the lexicographical ordering from the indexing of segments by left endpoints.  
\end{enumerate}
Step (1) applies the Bron-Kerbosch algorithm \cite{Bron1973,Birand2014} to find all maximal cardinality independent sets. The importance of the lexicographical ordering in step (5) will become apparent after Example \ref{ex:lex} below.

The output of this process is a vertex cover $P$. In the case that $\tau=0$, this vertex cover represents a pseudoknotted structure by Definition \ref{def:pseudoknot-segment}, where the number of pseudoknots is quantified by the minimum cardinalities of vertex covers of minimum weight. In the cases where $\tau> 0$, the minimum-weight minimum-cardinality vertex cover represents a $\tau$-pseudoknotted structure by Definition \ref{def:pseudoknot-tau}.

After identifying pseudoknots, one may still want to analyze the crossingless secondary structures of the RNA molecule. Therefore the last part of the process removes all chords which compose a pseudoknot, leaving a set of chords $C'=C-P$. In the case of $\tau = 0$, the set of chords $C'$ is crossingless. In the case $\tau > 0$, it may be that chords in $C'$ cross, however $C'$ comprises independent crossings and nestings, i.e. no two segments cross in the augmented segment partition of $C'$. Setting $\tau=\infty$ (which corresponds to an augmented segment partition), we enter the secondary structure classification subroutine (\texttt{classifyBases}) assessing secondary structure using the definitions given in Section \ref{subsec:secondary_structures}, with independent crossings handled as nestings (and thus a type of stem). 

This entire process is summarized in a flow chart (see Figure \ref{fig:flow_chart}) in the Appendix.

\begin{example}[Lexicographical Ordering Matters] \label{ex:lex}
    Here we apply the $\tau=0$ reduction method to the RNA structure 3DIG from the Protein Data Bank \cite{Serganov2008, PDB455} (see Figure \ref{fig:pdb455}). There are two choices for a minimum-cardinality minimum-weight vertex cover, here of cardinality three. Namely, both covers contain the two segments highlighted in red, and differ by whether the cover contains segment $\{(40,52)\}$ or segment $\{(41,54)\}$. Removing either cover yields a chord diagram with no crossings, and finding the secondary structures, we may obtain two different pseudoknot types depending on the cover removed. If we remove the cover containing segment $\{(40,52)\}$, then the pseudoknot corresponding to $\{(40,52)\}$ connects a bulge to a bulge. However if instead we remove the cover containing segment $\{(41,54)\}$ then the pseudoknot corresponding to $\{(41,54)\}$ connects an interior loop to another interior loop. As a result, the two choices for a vertex cover have a different effect on the secondary structure classification and consequently pseudoknot typing. 
\begin{figure}
    \centering
    \includegraphics[width=\textwidth]{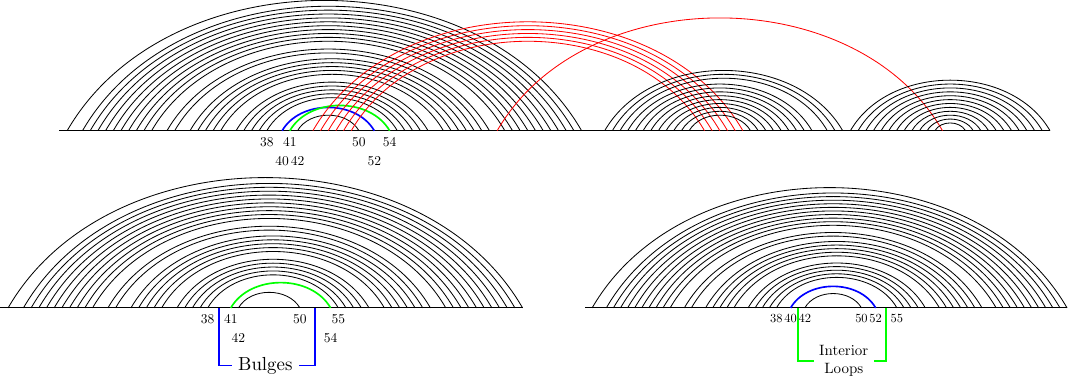}
    \caption{RNA structure with bpRNA reference name PDB\_455 and PDB reference name 3DIG. Structures such as nestings not relevant for the discussion have been omitted for clarity.}
    \label{fig:pdb455}
\end{figure}

\end{example}

\section{Discussion}
\label{sec:discussion}

We applied the $\tau$-Segment Graph Method with $\tau=0$ to independently verify the quantities reported in bpRNA-1m(90) \cite{bpRNA}. By Definition \ref{def:pseudoknot-segment}, structures are pseudoknotted when their segment graphs contain at least one edge, and the number of pseudoknots is quantified by the minimum cardinalities of vertex covers of minimum weight. In agreement with \cite{bpRNA}, we obtained 3,320 graphs containing at least one edge in $G_\cS$ from RNA structures and a total quantity of 7,164 pseudoknots, as determined by the sum over the cardinalities of the vertex covers.
Applying the $\tau$-Segment Graph Method with $\tau=\infty$  (the augmented segment graph method) for every structure in bpRNA-1m(90), we found found a minimum vertex cover for each structure. The number of pseudoknots (the sum of vertex cover numbers over all graphs) was 6,548 with this method.

With $\tau=0$, we found 31 unique RNA structures containing 13 or more pseudoknots. These structures are listed in Table \ref{tab:mostPK}. 
When the method was applied with $\tau=\infty$, we found that the same 31 structures contained the most pseudoknots amongst all structures in the database. 
With the exception of the last structure, \textit{Oceanobacillus iheyensis}, from $\tau=0$ to $\tau=\infty$ there was a uniform decrease in pseudoknotting by two that resulted from a single 3-crossing being consolidated into one segment by $\tau$-reduction. 
This uniformity in behavior is explained by the fact that all but the last structure are of type 23S prokaryotic ribosomal RNA, originating in various bacterial organisms \cite{MolecularBiologyofRNA}.

Over the entire bpRNA-1m(90) database, a total of 573 structures had a decrease in numbers of pseudoknots when analyzed with the $\tau=\infty$ versus $\tau=0$ methods. Of these, 531 structures decreased in quantity of pseudoknots by 1, 41 structures by 2, and 1 structure decreased by 3 (bpRNA\_CRW\_55316, \textit{Plasmodium falciparum}). Of the 41 structures which decreased by 2, there were 6 unique RNA types with 36 of them being of type 23S. 
Structures that changed from having a nonzero quantity of pseudoknots to zero pseudoknots are shown in Table \ref{tab:zeroPK}. Of these, one structure (\emph{Homo sapiens}) decreased from 2 to 0 pseudoknots. All other structures decreased from 1 to 0 pseudoknots.

\subsection{Maximum Values of \texorpdfstring{$\tau$}{tau} and Persistence of Partitions}
Let $\tau \ge 1$. As distances between chords are finite, there is a minimum value of $\tau$, say $\tau_{m}$, such that for any $\tau_{\ast} \ge \tau_m$ the $\tau_{\ast}$-segment partition and the $\tau_m$-segment partition are identical. The quantity $\tau_m$ is precisely the minimum value of $\tau$ such that the $\tau$-segment partition is equivalent to the augmented segment partition. For all structures in bpRNA-1m(90), we calculate $\tau_m$ by first finding the augmented segment partition, and then finding the $\tau$-segment partition for each $\tau > 0$ until the $\tau$-segment partition is equal to the augmented segment partition. We find that the average $\tau_m$ is $13.035$ and the median is $8$. The mean absolute deviation is $10.96$ and the median absolute deviation is $2$. There are $323$ structures with $\tau_m$ at least $17$, and $33$ structures with $\tau_m$ at least $100$.

Structures with large $\tau_m$ contain correspondingly large bulges and internal loops; large $\tau_m$ results from large gaps between chords which are nested but not $\tau$-near for many values of $\tau$. See for example Figure \ref{fig:largeTM}. We verified this by keeping track of $\tau$-segment partitions during the process of calculating $\tau_m$. In sum, persistent $\tau$-segment partitions are indicative of large bulges and internal loops.

\subsection{Classifying Bases}

We implemented the \texttt{classifyBases} routine with $\tau=0$ to analyze secondary structures and compare quantities obtained from the bpRNA-1m database (see also \cite[Figure 8b]{bpRNA}). Quantities are shown in Table \ref{fig:PK_types_table} (left).
We observed slight discrepancies in pseudoknot type counts, though the general shape of the distribution is the same. 
The discrepancies with the $\tau=0$ method arise from the labeling of multiloops and external loops, as some structures in \cite{bpRNA} have bases in the external loop that are labeled as a multiloop base. This is a bpRNA software bug that has since been fixed in a fork of \cite{bpRNArepo}. The structure bpRNA\_CRW\_10025 is one example in which the secondary structure labeling is incorrect; bases 357, 385-393, 433-440, 604-633, 690-695, and 728-742 are labeled as part of a multiloop in bpRNA-1m, but by our definition they are part of the exterior loop. Note that the counts of pseudoknot types in Figure \ref{fig:PK_types_table} (left) differ only when an exterior loop or multiloop is part of the pseudoknot type. Figure \ref{fig:PK_types_table} (right) shows a comparison of pseudoknot type counts between the $\tau=0$ method and the $\tau=\infty$ method.

\subsection{Calculation of Genus and Clique Numbers}\label{sec:genus}
After implementing the $\tau$-segment graph method with $\tau=0$ and $\tau=\infty$ we calculated the genus and maximum clique numbers of $D_\cS$ and $D_\cA$ and the segment graphs $G_\cS$ and $G_\cA$ respectively in \MATLAB. The results are reported in Figure \ref{fig:genus_histogram}, Table \ref{tab:clique_numbers}, and Figure \ref{fig:genus_bubble}.
Out of 3,320 segment graphs, 3,208 are forests, and out of 3,320 augmented segment graphs, 3,210 are forests. Table \ref{tab:treeorder} shows the frequency of forests with a given maximum tree size. This is important to note in the context of using genus and vertex covers for pseudoknot quantification. By Theorem \ref{thm:genusvertexcover}, if the intersection graph of a chord diagram $D$ is a forest $F$, then the genus $\gamma(D)$ is equal to $\beta(F)$. That is, for acyclic intersection graphs, an increase in genus implies an increase in the vertex cover. From this, we see that the genus of a corresponding chord diagram of an RNA structure is a robust quantifier of pseudoknot complexity. This is further supported by the bubble charts in Figure \ref{fig:genus_bubble}.

\section*{Acknowledgements}

The authors were partially supported by the National Science Foundation, DMS–2204148 and The Thomas F. and Kate Miller Jeffress Memorial Trust,
Bank of America, Trustee.

\Addresses

\clearpage

\section*{Appendix}


\begin{figure}[hb]
    \centering    
    \includegraphics[width=1\textwidth]{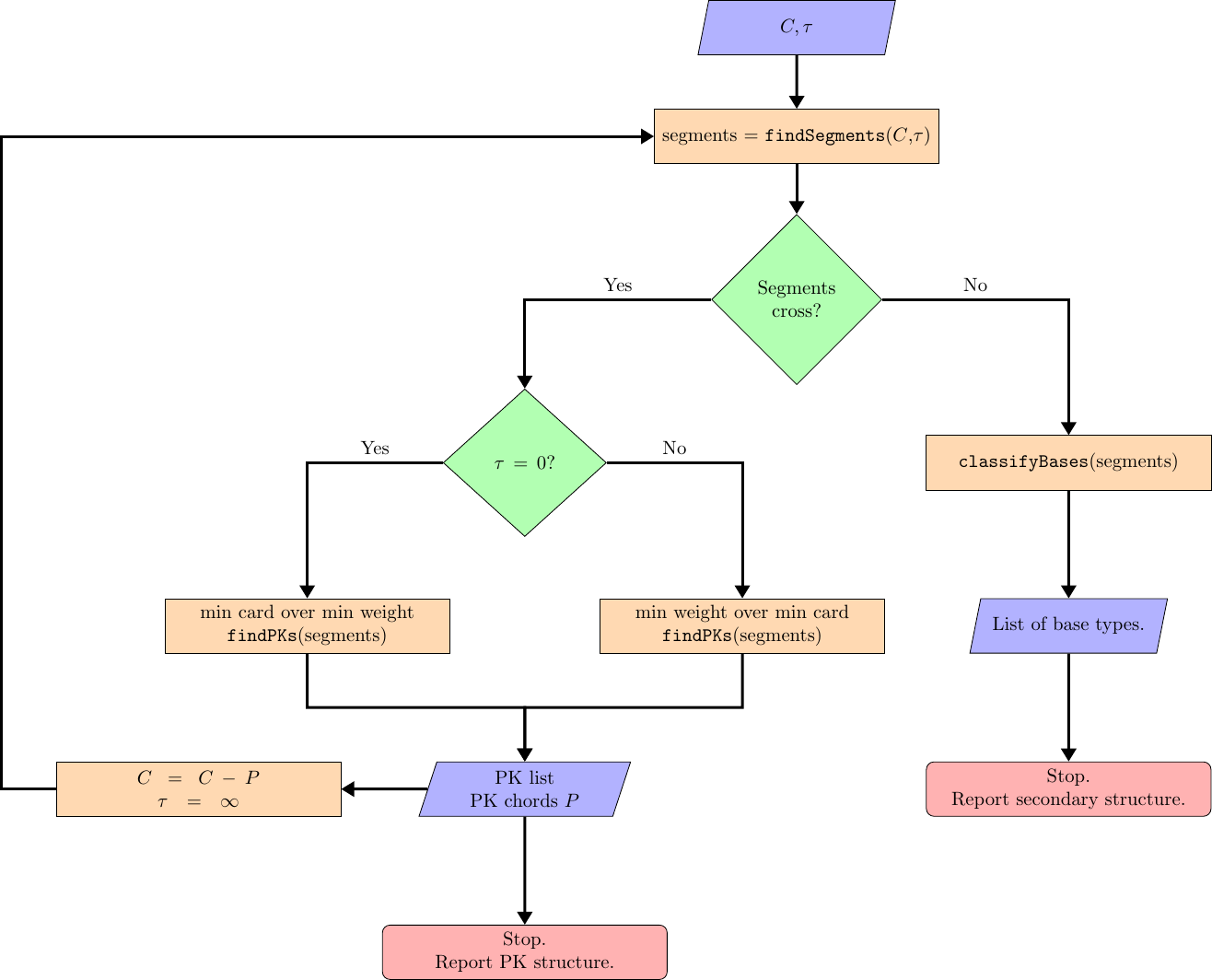}
    \caption{A flow chart summarizing the the pseudoknot and secondary structure identification and quantification process described in Section \ref{process}.}
    \label{fig:flow_chart}
\end{figure}


\begin{figure}[ht]
    \centering    
    \includegraphics[width=1\textwidth]{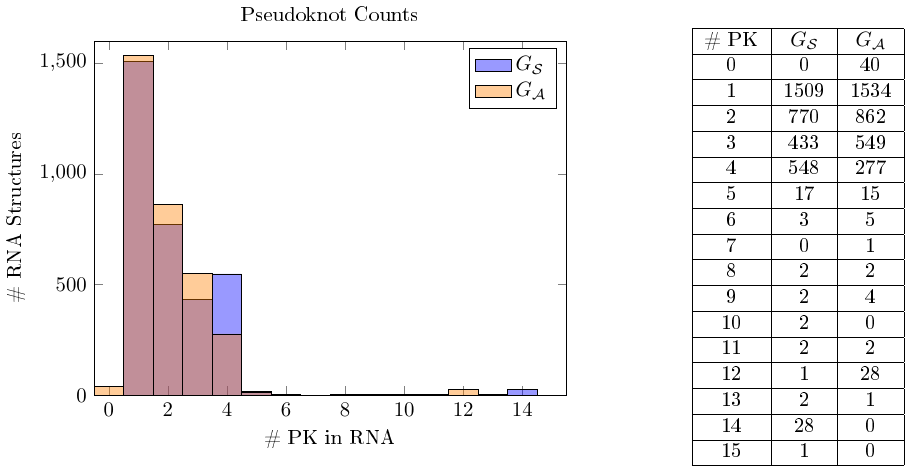}
    \caption{A comparison of total number of pseudoknots between the segment graph method ($\tau=0$) and augmented segment graph method ($\tau=\infty$.) For each method, the number of structures containing $r$ pseudoknots is given.}
    \label{fig:numPKhist}
\end{figure}

\begin{table}[ht]
    \centering
    \resizebox{\textwidth}{!}{
\begin{tabular}[t]{lllll}
ID & Domain & Organism & $G_{\mathcal{S}}$ & $G_{\mathcal{A}}$ \\ 
\hline 
CRW\_55315 & Eukaryota & Euglena gracilis & 15 & 13 \\ 
CRW\_55268 & Bacteria & Acinetobacter calcoaceticus & 14 & 12 \\ 
CRW\_55269 & Bacteria & Aeromonas hydrophila & 14 & 12 \\ 
CRW\_55271 & Bacteria & Bartonella bacilliformis & 14 & 12 \\ 
CRW\_55275 & Bacteria & Burkholderia mallei & 14 & 12 \\ 
CRW\_55276 & Bacteria & Bordetella pertussis & 14 & 12 \\ 
CRW\_55279 & Bacteria & Clostridium botulinum B & 14 & 12 \\ 
CRW\_55283 & Bacteria & Citrobacter freundii & 14 & 12 \\ 
CRW\_55284 & Bacteria & Campylobacter jejuni & 14 & 12 \\ 
CRW\_55285 & Bacteria & Chlamydophila psittaci 6BC & 14 & 12 \\ 
CRW\_55287 & Bacteria & Deinococcus radiodurans & 14 & 12 \\ 
CRW\_55290 & Bacteria & Enterococcus faecalis & 14 & 12 \\ 
CRW\_55291 & Bacteria & Erysipelothrix rhusiopathiae (str. 715) & 14 & 12 \\ 
CRW\_55292 & Bacteria & Haemophilus influenzae (operons A-F) & 14 & 12 \\ 
CRW\_55295 & Bacteria & Leptospira interrogans & 14 & 12 \\ 
CRW\_55296 & Bacteria & Lactococcus lactis & 14 & 12 \\ 
\hline 
\end{tabular}
\begin{tabular}[t]{lllll}
ID & Domain & Organism & $G_{\mathcal{S}}$ & $G_{\mathcal{A}}$ \\ 
\hline 
\multicolumn{5}{ c }{$\vdots$} \\
CRW\_55297 & Bacteria & Listeria monocytogenes & 14 & 12 \\ 
CRW\_55298 & Bacteria & Listeria monocytogenes & 14 & 12 \\ 
CRW\_55299 & Bacteria & Mycoplasma genitalium & 14 & 12 \\ 
CRW\_55303 & Bacteria & Neisseria gonorrhoeae & 14 & 12 \\ 
CRW\_55305 & Bacteria & Pseudomonas aeruginosa & 14 & 12 \\ 
CRW\_55306 & Bacteria & Plesiomonas shigelloides & 14 & 12 \\ 
CRW\_55307 & Bacteria & Ruminobacter amylophilus & 14 & 12 \\ 
CRW\_55308 & Bacteria & Rickettsia prowazekii (str. Madrid E) & 14 & 12 \\ 
CRW\_55312 & Bacteria & Staphylococcus carnosus & 14 & 12 \\ 
CRW\_55313 & Bacteria & Thermotoga maritima & 14 & 12 \\ 
CRW\_55314 & Eukaryota & Chlamydomonas reinhardtii & 14 & 12 \\ 
CRW\_55317 & Eukaryota & Spinacia oleracea & 14 & 12 \\ 
CRW\_55338 & Eukaryota & Cyanophora paradoxa & 14 & 12 \\ 
CRW\_55270 & Bacteria & Bacillus anthracis & 13 & 11 \\ 
PDB\_647 & Bacteria & Oceanobacillus iheyensis & 13 & 11 \\ 
\hline 
\end{tabular}
    }
    \caption{The $31$ RNA structures with at least 13 pseudoknots when analyzed with the segment graph method. The rightmost column compares the number of pseudoknots in each structure via the $\tau$-segment graph method with $\tau = \infty$. All structures with the exception of PDB\_647 are RNA type 23S ribosomal RNA.}
    \label{tab:mostPK}
\end{table}

\begin{table}[ht]
    \centering
    \resizebox{\textwidth}{!}{
\begin{tabular}[t]{lllll}
ID & Domain & Organism & Length & Method \\ 
\hline 
CRW\_1213 & Bacteria & Actinomyces israelii & 734 & CSA \\ 
CRW\_1219 & Bacteria & Actinomyces israelii & 1145 & CSA \\ 
CRW\_1563 & Bacteria & Clavibacter sp. R1\_2\_cr & 476 & CSA \\ 
CRW\_1725 & Bacteria & Arthrobacter sp. & 300 & CSA \\ 
CRW\_17723 & Bacteria & Lachnospira multipara & 977 & CSA \\ 
CRW\_17729 & Bacteria & Moorella thermoautotrophica & 869 & CSA \\ 
CRW\_17730 & Bacteria & Moorella thermoautotrophica & 821 & CSA \\ 
CRW\_17811 & Bacteria & Thermoanaerobacter acetoethylicus & 770 & CSA \\ 
CRW\_17823 & Bacteria & Thermoanaerobacter ethanolicus & 650 & CSA \\ 
CRW\_17834 & Bacteria & Thermoanaerobacterium thermosulfurigenes & 930 & CSA \\ 
CRW\_20267 & Bacteria & Marigold phyllody phytoplasma & 1015 & CSA \\ 
CRW\_20554 & Bacteria & Mycoplasma collis & 372 & CSA \\ 
CRW\_20606 & Bacteria & Beet leafhopper transmitted virescence phytoplasma & 700 & CSA \\ 
CRW\_20626 & Bacteria & Potato witches'-broom phytoplasma & 658 & CSA \\ 
CRW\_20629 & Bacteria & Paulownia witches'-broom phytoplasma & 698 & CSA \\ 
CRW\_3719 & Bacteria & Actinomycetales sp. & 472 & CSA \\ 
CRW\_3726 & Bacteria & Actinomycetales sp. & 503 & CSA \\ 
CRW\_3729 & Bacteria & Actinomycetales sp. & 502 & CSA \\ 
CRW\_3732 & Bacteria & Actinomycetales sp. & 478 & CSA \\ 
CRW\_4109 & Bacteria & Mycobacterium xenopi & 942 & CSA \\ 
CRW\_4363 & Bacteria & Streptomyces abikoensis & 1177 & CSA \\ 
\hline 
\end{tabular}
\begin{tabular}[t]{lllll}
ID & Domain & Organism & Length & Method \\ 
\hline 
\multicolumn{5}{c}{$\vdots$}\\
CRW\_4401 & Bacteria & Streptomyces mobaraensis & 1197 & CSA \\ 
CRW\_4409 & Bacteria & Streptomyces olivoreticuli & 1216 & CSA \\ 
CRW\_4416 & Bacteria & Streptomyces salmonis & 1136 & CSA \\ 
CRW\_4449 & Bacteria & coryneform actinomycete B755 & 679 & CSA \\ 
CRW\_4908 & Bacteria & Acidocella facilis & 922 & CSA \\ 
CRW\_4910 & Bacteria & Acidiphilium angustum & 977 & CSA \\ 
CRW\_4918 & Bacteria & Acidiphilium sp. & 944 & CSA \\ 
CRW\_7455 & Bacteria & unidentified eubacterium 37SW-1 & 277 & CSA \\ 
CRW\_7488 & Bacteria & Proteobacteria sp & 484 & CSA \\ 
CRW\_7494 & Bacteria & uncultured alpha proteobacterium & 410 & CSA \\ 
CRW\_7502 & Bacteria & uncultured alpha proteobacterium & 222 & CSA \\ 
CRW\_7614 & Bacteria & Nitrobacter sp. & 452 & CSA \\ 
CRW\_7802 & Bacteria & Rhodovulum euryhalinum & 1138 & CSA \\ 
CRW\_7938 & Bacteria & Sphingomonas asaccharolytica & 629 & CSA \\ 
CRW\_8046 & Bacteria & uncultured alpha proteobacterium & 751 & CSA \\ 
CRW\_8048 & Bacteria & uncultured alpha proteobacterium & 730 & CSA \\ 
CRW\_8050 & Bacteria & uncultured alpha proteobacterium & 690 & CSA \\ 
PDB\_567 & artificial sequences & synthetic construct & 35 & X-RAY \\ 
PDB\_512 & Eukaryota & Homo sapiens & 12 & X-RAY \\ 
\hline 
\end{tabular}
    }
    \caption{The 40 RNA structures with nonzero quantity of pseudoknots when analyzed with the segment graph method but zero pseudoknots using $\tau=\infty$ segment graph method. One structure (\textit{Homo sapiens}) decreased from 2 to 0 pseudoknots. All  other structures decreased from 1 to 0 pseudoknots.}
    \label{tab:zeroPK}
\end{table}

\begin{figure}[ht]
    \centering
    \includegraphics[width=\textwidth]{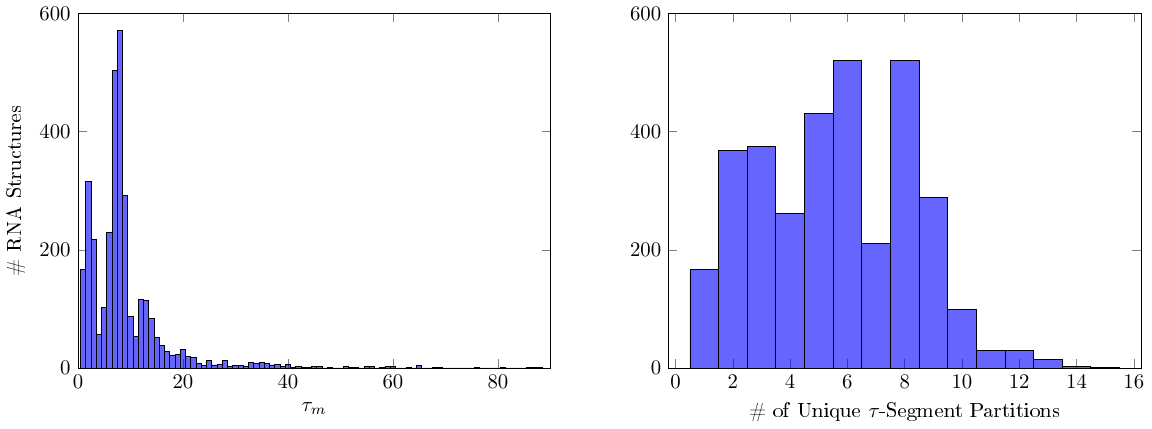}
    \caption{Left: Distribution of $\tau_m$ over all RNA structures in bpRNA-1m(90) restricted to values of $\tau_m$ within one and a half standard deviations from the mean. Right: Distribution of number of unique $\tau$-segment partitions.}
    \label{fig:taumax_histogram}
\end{figure}

\begin{figure}[ht]
    \centering
    \includegraphics[width=0.9\textwidth]{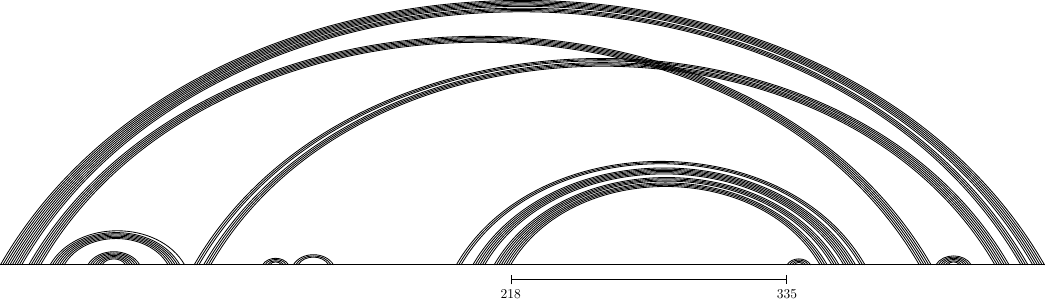}
    \caption{The RNA structure bpRNA\_RFAM\_4761 is an example of a structure with large $\tau_m=117$ resulting from an interior loop.}
    \label{fig:largeTM}
\end{figure}


\begin{figure}[h]
\centering
\begin{footnotesize}
\begin{tabular}{| c | c | c |}
\hline
PK type & bpRNA-1m & $G_{\mathcal{S}}$\\ \hline
E-E & 0 & 0 \\ \hline
E-X & 0 & 0 \\ \hline
X-X & 0 & 0 \\ \hline
B-B & 6 & 6 \\ \hline
I-I & 9 & 9 \\ \hline
B-E & 10 & 10 \\ \hline
I-M & 49 & 48 \\ \hline
E-M & 53 & 53 \\ \hline
I-X & 57 & 58 \\ \hline
E-I & 64 & 64 \\ \hline
M-X & 100 & 104 \\ \hline
B-I & 153 & 153 \\ \hline
B-X & 158 & 169 \\ \hline
H-I & 194 & 194 \\ \hline
H-H & 261 & 261 \\ \hline
B-M & 377 & 366 \\ \hline
E-H & 588 & 588 \\ \hline
H-X & 670 & 847 \\ \hline
M-M & 711 & 707 \\ \hline
B-H & 1826 & 1826\\ \hline
H-M & 1878 & 1701\\ \hline
\end{tabular}
\hspace{3cm}
\begin{tabular}{| c | c | c |}
\hline
PK type & $G_\mathcal{S}$ & $G_{\mathcal{A}}$\\ \hline
E-E & 0 & 0 \\ \hline
E-X & 0 & 0 \\ \hline
X-X & 0 & 0 \\ \hline
B-B & 6 & 5 \\ \hline
I-I & 9 & 7 \\ \hline
B-E & 10 & 8 \\ \hline
I-M & 48 & 48 \\ \hline
E-M & 53 & 54 \\ \hline
I-X & 58 & 60 \\ \hline
E-I & 64 & 57 \\ \hline
M-X & 104 & 102 \\ \hline
B-I & 153 & 82 \\ \hline
B-X & 169 & 5 \\ \hline
H-I & 194 & 196 \\ \hline
H-H & 261 & 258 \\ \hline
B-M & 366 & 9 \\ \hline
E-H & 588 & 584 \\ \hline
M-M & 707 & 705 \\ \hline
H-X & 847 & 845 \\ \hline
H-M & 1701 & 1699\\ \hline
B-H & 1826 & 1824\\ \hline
\end{tabular}
\end{footnotesize}
     \caption{(Left) A comparison of counts of pseudoknot types reported in bpRNA-1m  versus our $G_{\cS}$ method with $\tau=0$. Discrepancies result from a known bug in the bpRNA software. (Right) A comparison of counts of pseudoknot types using our $G_{\cS}$ and $G_{\cA}$ methods with $\tau=0$ and $\tau=\infty$, respectively.} \label{fig:PK_types_table}
\end{figure}


\begin{figure}[ht]
    \centering
    \includegraphics[width=\textwidth]{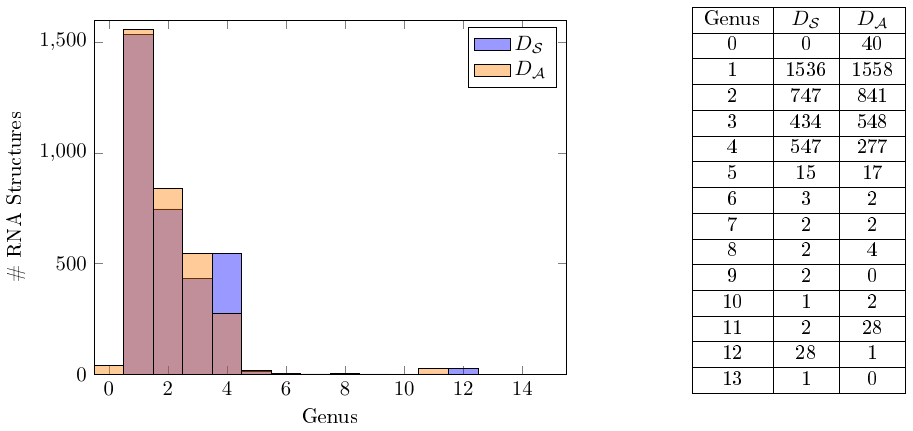}
    \caption{A comparison of the genus of the segment graph and the augmented segment graph.}
    \label{fig:genus_histogram}
\end{figure}

\begin{table}[ht]
    \centering
    \begin{tabular}{| c | c | c |}
    \hline
       $\omega(G)$  & $G_{\cS}$ & $G_{\cA}$\\ \hline
        1 & 0 &  40    \\ \hline
        2 & 3214 & 3177  \\ \hline
        3 & 60 &  96  \\ \hline
        4 & 46 &   7  \\ \hline
    \end{tabular}
    \caption{The frequency of clique numbers for segment and augmented segment graphs.}
    \label{tab:clique_numbers}
\end{table}

\begin{table}[ht]
    \centering
    \resizebox{0.6\textwidth}{!}{
    \begin{tabular}{| c | c | c |}
    \hline
        Max Tree Order & $G_{\cS}$ & $G_{\cA}$ \\ 
        \hline 
        1 & 0 & 40 \\ \hline
        2 & 1449 & 1410 \\ \hline
        3 & 998 & 999 \\ \hline
        4 & 637 & 637 \\ \hline
        5 & 27 & 27 \\ \hline
        6 & 71 & 71 \\ \hline
    \end{tabular}\hspace{1cm}
    \begin{tabular}{| c | c | c |}
    \hline
        Max Tree Order & $G_{\cS}$ & $G_{\cA}$ \\ 
        \hline 
        7 & 6 & 6 \\ \hline
        8 & 7 & 7 \\ \hline
        9 & 7 & 7 \\ \hline
        10 & 1 & 1 \\ \hline
        11 & 3 & 3 \\ \hline
        12 & 2 & 2 \\ \hline
    \end{tabular}
    
    }
    \caption{For each segment and augmented segment graph that is a forest, we calculate the maximum order of a tree component. 
    }
    \label{tab:treeorder}
\end{table}

\begin{figure}[ht]
    \centering
    \includegraphics[width=1.0\textwidth]{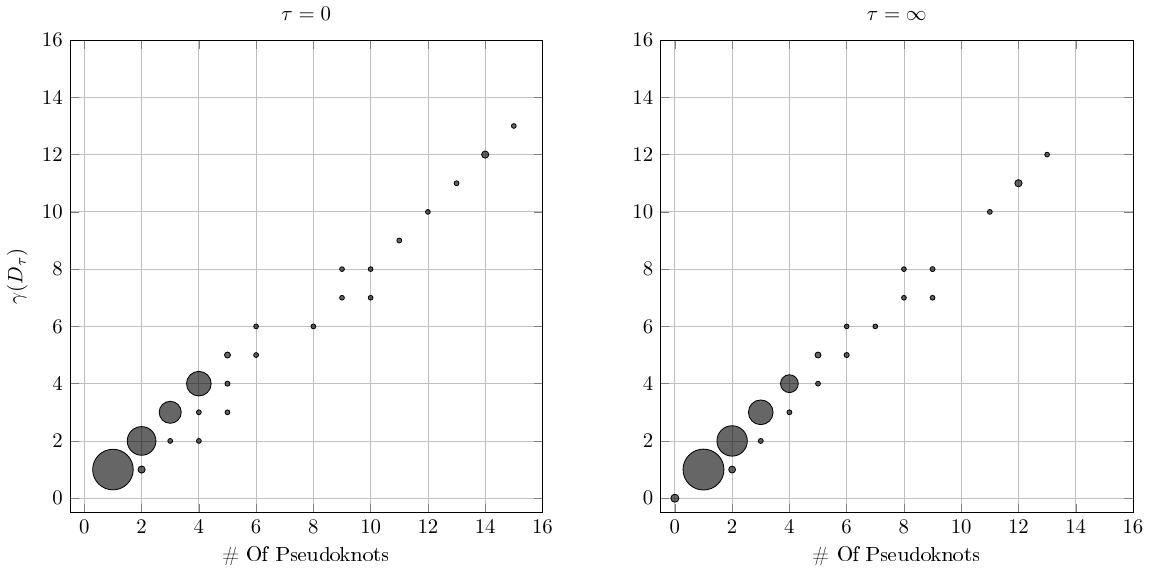}
    \caption{Two bubble chart comparisons. Left: Comparison between genera of the chord diagrams $D_\cS$ and pseudoknot count via the $\tau=0$ method. Right: Comparison between genera of the chord diagrams $D_\cA$ and pseudoknot count via the $\tau=\infty$ method.}
    \label{fig:genus_bubble}
\end{figure}

\clearpage

\bibliographystyle{plain}
\bibliography{references.bib}

\end{document}